\documentclass[twocolumn]{aastex631} 

\usepackage[]{mathtools}
\usepackage{bm}
\usepackage{graphicx}
\usepackage{subfigure} 
\usepackage{float}
\usepackage{verbatim}
\usepackage{booktabs}
\usepackage{multirow}
\usepackage{color}
\usepackage{mathtools}
\usepackage{longtable}
\usepackage{comment}
\usepackage{amssymb}
\usepackage{bm}
\usepackage[dvipsnames, table,xcdraw]{xcolor}
\usepackage[utf8]{inputenc}

\newcommand{\qh}[1]{\textcolor{red}{#1}}

\newcommand{\Msun}{\ensuremath{M_\odot}}

\newcommand{\lbf}{\log_{10} \mathcal{B}^{\mathcal{OS}}_{\mathcal{S}}}
\newcommand{\bayesf}{\mathcal{B}^{\mathcal{OS}}_{\mathcal{S}}}

\begin{document}

\title{GW231123: Overlapping Gravitational Wave Signals?}

\author[0000-0002-3033-6491]{Qian Hu}
\email{Qian.Hu@glasgow.ac.uk}
\affiliation{Institute for Gravitational Research, School of Physics and Astronomy, University of Glasgow, Glasgow, G12 8QQ, United Kingdom}

\author[0000-0001-9161-7919]{Harsh Narola}
\email{h.b.narola@uu.nl}
\affiliation{Institute for Gravitational and Subatomic Physics (GRASP), Utrecht University, Princetonplein 1, 3584 CC Utrecht, The Netherlands}
\affiliation{Nikhef---National Institute for Subatomic Physics, Science Park 105, 1098 XG Amsterdam, The Netherlands}

\author[0009-0007-5496-2159]{Jef Heynen}
\affiliation{Centre for Cosmology, Particle Physics and Phenomenology - CP3, Universit\'e Catholique de Louvain, Louvain-La-Neuve, B-1348, Belgium}

\author[0000-0003-1829-7482]{Mick Wright}
\affiliation{Institute for Gravitational and Subatomic Physics (GRASP), Utrecht University, Princetonplein 1, 3584 CC Utrecht, The Netherlands}
\affiliation{Nikhef---National Institute for Subatomic Physics, Science Park 105, 1098 XG Amsterdam, The Netherlands}

\author[0000-0002-6508-0713]{John Veitch} 
\affiliation{Institute for Gravitational Research, School of Physics and Astronomy, University of Glasgow, Glasgow, G12 8QQ, United Kingdom}

\author[0000-0003-2888-7152]{Justin Janquart}
\affiliation{Centre for Cosmology, Particle Physics and Phenomenology - CP3, Universit\'e Catholique de Louvain, Louvain-La-Neuve, B-1348, Belgium}
\affiliation{Royal Observatory of Belgium, Avenue Circulaire, 3, 1180 Uccle, Belgium}

\author[0000-0001-6800-4006]{Chris Van Den Broeck}
\affiliation{Institute for Gravitational and Subatomic Physics (GRASP), Utrecht University, Princetonplein 1, 3584 CC Utrecht, The Netherlands}
\affiliation{Nikhef---National Institute for Subatomic Physics, Science Park 105, 1098 XG Amsterdam, The Netherlands}

\date{\today}

\begin{abstract}
    The recently discovered gravitational wave event GW231123 was interpreted as the merger of two black holes with a total mass of 190-265\,\Msun, making it the heaviest such merger detected to date. Whilst much of the post-discovery literature has focused on its  astrophysical origins, primary analyses have exhibited considerable discrepancies in the measurement of source properties between waveform models, which cannot reliably be reproduced by simulations. Such discrepancies may arise when an unaccounted overlapping signal is present in the data, or from phenomena that produce similar effects, such as gravitational lensing or overlapping noise artifacts. In this work, we analyse GW231123 using a flexible model that allows for two overlapping signals, and find that it is favoured over the isolated signal model with Bayes factors of $\sim 10^2 - 10^{4}$, depending on the waveform model. These values lie within the top few per cent of the background distribution. Similar effects are not observed in GW190521, another high-mass event. Under the overlapping signals model, discrepancies in the measurement of source properties between waveform models are largely mitigated. We also find that neglecting an additional signal in overlapping-signal data can lead to discrepancies in the estimated source properties resembling those reported in GW231123. {Although the overlapping signal model provides a higher Bayesian evidence, the astrophysical prior probability of two short signals overlapping is low. However, we find that the two recovered sources show similar properties. This, taken with the higher evidence of the two signal model, suggests that gravitational lensing may provide an alternative explanation. }
\end{abstract}

%

\section{\label{sec1}Introduction}The LIGO--Virgo--KAGRA collaboration~\citep{LIGOScientific:2014pky, VIRGO:2014yos, KAGRA:2020tym} recently reported the detection of the gravitational wave (GW) signal, GW231123\_135430 (hereafter referred to as GW231123), from the merger of a binary black hole (BBH) system~\citep{LIGOScientific:2025rsn}. With an estimated total mass of 190-265 \Msun, it became the heaviest BBH system detected to date, surpassing GW190521~\citep{LIGOScientific:2020iuh}, whose total mass was 126-170 \Msun. The individual black hole masses of GW231123 lie within or above the pair-instability black hole mass gap predicted by standard stellar evolution theory~\citep{Farmer:2019jed, Farmer:2020xne}, making the event particularly intriguing from an astrophysical perspective. Several formation scenarios have been proposed, including hierarchical mergers~\citep{Stegmann:2025cja, Li:2025fnf, Li:2025pyo}, mergers in active galactic nucleus disks~\citep{Bartos:2025pkv, Delfavero:2025lup},  primordial black holes~\citep{Yuan:2025avq, DeLuca:2025fln}, cosmic strings~\citep{Cuceu:2025fzi}, and evolution of massive stars under specific conditions such as low-metallicity~\citep{Gottlieb:2025ugy, Tanikawa:2025fxw}, moderate magnetic fields~\citep{Gottlieb:2025ugy}, and high spins~\citep{Croon:2025gol, Popa:2025dpz}. Constraints on fundamental physics are also made based on the parameter estimation of GW231123~\citep{Caputo:2025oap, Aswathi:2025nxa, Wang:2025rvn}. 

Determining the formation channel of a BBH depends on its source properties, i.e., masses, distance, spins, and such. Measuring these properties for an unusual system such as GW231123 is also a challenging task which heavily relies on how accurately we can model the signal. In this scenario, these are known as waveform models which describe the evolution of spacetime~\citep{Pratten:2020ceb, Varma:2019csw, Estelles:2021gvs, Thompson:2023ase}. For GW231123, the measured source properties using different waveform models show significant discrepancies~\citep{LIGOScientific:2025rsn}, which cannot be reliably reproduced by the simulations.
The scale of discrepancies appears to be larger compared to the expected statistical fluctuations and also larger compared to the systematic differences between the waveform models. This behaviour may be attributed to, but not limited to, inaccuracy of the waveform models, missing physical effects, or noise fluctuations in the data~\citep{Ray:2025rtt, Siegel:2025xgb, Romero-Shaw:2022fbf}.

An additional source for such discrepancies may be the presence of overlapping signals~\citep{Pizzati:2021apa, Himemoto:2021ukb, Samajdar:2021egv, Relton:2021cax,  Hu:2022bji, Janquart:2022fzz, Wang:2023ldq, Johnson:2024foj, Baka:2025yqx}, i.e., signals arriving at the detectors close to each other in time. Since GW detection rate estimates disfavour such scenarios~\citep{LIGOScientific:2025pvj}, the state-of-the-art parameter estimation pipelines do not account for overlapping signals~\citep{Ashton:2018jfp}. However, when not accounted for, they may bias the measurements of source properties such as masses, distance, and spins~\citep{Samajdar:2021egv, Relton:2021cax, Hu:2025vlp}. The fainter overlapping signal may be fitted differently by different waveforms, leading to waveform systematics which cannot be reproduced in simulations accounting for only one isolated signal. This seems to be the case for GW231123, i.e., the estimated source properties using different waveform models do not agree, raising the question: \textit{could there be an additional signal overlapping with the GW231123?} To address this question, we analyse GW231123 with a flexible signal model that accounts for two independent overlapping signals, {and find that it is supported by the Bayes factor against the single signal model. }

In principle, such a model could also fit other additional power in the data that overlaps with the signal, such as faint noise artefacts~\citep{Ray:2025rtt}. 
More importantly, significant interest in determining the presence of an additional signal near GW231123 is generated by the recent gravitational lensing interpretation of the event~\citep{xikai:2025, goyalab:2025, Liuab:2025, LIGOScientific:2025cwb}. Specifically, the scenario where a gravitational wave signal is lensed by a massive object in its path~\citep{Takahashi:2003ix}, and the two images produced by lensing are not well-separated in time, they can imitate the features of two overlapping signals~\citep{Liu:2023ikc}. This lensing scenario may be viewed as a specific sub-case of our overlapping signal case of more generalised overlapping signal analysis, and may affect the significance of its lensing interpretation~\citep{Rao:2025poe}. To make a confident detection of a lensed gravitational wave signal, it would be crucial to rule out the possibility of coincidental overlap of signals. {We calculate the prior odds of astrophysical overlapping signals in the GW231123 scenario, and find it rules out this possibility. In addition, we find the two recovered sources under the overlapping signal model have similar properties and sky locations. These results point to a potential favour towards the lensing interpretation. }

\section{\label{sec:modsel} Bayesian model selection}We employ the Bayesian model selection technique to determine which model, two overlapping signals or the isolated signal, fits better to the data corresponding to GW231123. Specifically, we compute the Bayesian evidence $\mathcal{Z}$,
\begin{equation}\label{eq:evidence}
\mathcal{Z} = \int \mathrm{d}\vec{\theta}~\pi(\vec{\theta})~\mathcal{L}(\vec{d} | \vec{\theta}),
\end{equation}
for each model. In Eq.~\eqref{eq:evidence}, $\vec{\theta}$ denotes the model parameters, $\vec{d}$ denotes the detector strain, $\pi(\vec{\theta})$ denotes the prior on $\vec{\theta}$ and $\mathcal{L}(\vec{d} | \vec{\theta})$ denotes the likelihood of $\vec{d}$ given $\vec{\theta}$. For stationary Gaussian noise, the log-likelihood function up to a constant for $N$ signals can be written as~\citep{Veitch:2014wba, Samajdar:2021egv},
\begin{equation}
    \label{eq:likelihood}
    \ln{\mathcal{L}(\vec{d} | \vec{\theta})}  = -\frac{1}{2}\left<\vec{d}-\sum_{i=1}^N \vec{h}_i(\vec{\theta}) \right| \left. \vec{d}-\sum_{i=1}^N \vec{h}_i(\vec{\theta})\right>,
\end{equation}
where $<\cdot|\cdot>$ denotes the noise-weighted inner product~\citep{Finn:1992wt} and $\vec{h}(\vec{\theta})$ denotes the signal model.

The detector strains and the noise power spectral densities were obtained from the publicly available data release corresponding to GW231123~\citep{ligo_scientific_collaboration_2025_16004263}. We analyse a 4 seconds long segment centred at the trigger time of the event. The frequency range of our analysis is $[20,448]$ Hz. We adopt the prior distributions from the public data release~\citep{ligo_scientific_collaboration_2025_16004263}, with the exceptions that the chirp mass and luminosity distance priors are extended to 250 \Msun ~and 12 Gpc respectively. We restrict our analysis to two overlapping signals owing to the computation cost and the lower odds of more than two signals overlapping. When using the overlapping signals model, the prior on each signal is identical to the isolated signal model. Hence, the prior volume for the overlapping signals model is the square of the one for the isolated signal. We do not include detector calibration uncertainties since calibration envelopes are mostly consistent with zero~\citep{LIGOScientific:2025snk}. We use the Bilby codebase (v2.6.0) in combination with the Dynesty sampler (v2.1.5) to compute the evidence~\citep{Ashton:2018jfp, Speagle_2020}. 

\begin{table}[H]\centering
\renewcommand{\arraystretch}{1.2}
\begin{tabular}{|c|c|c|c|c|}
             \hline
             & \textbf{XPHM} & \textbf{NRSur} & \textbf{TPHM} & \textbf{XO4a} \\ \hline
             $\log_{10} \mathcal{Z}_{\mathcal{S}}$ & $-3774.14$ & $-3771.68$  & $-3769.88$ & $-3767.95$\\ \hline
             $\log_{10} \mathcal{Z}_{\mathcal{OS}}$ & $-3769.92$ & $-3769.72$  & $-3768.21$ & $-3767.74$\\ \hline
            $\lbf$ & 4.22 & 1.96 & 0.79 & 0.21 \\ \hline
\end{tabular}
\caption{Bayesian evidence obtained for each waveform when analysing GW231123 using the isolated signal model (first row) and the overlapping signals model (second row). 
    The third row shows log Bayes factors comparing the overlapping signals model to the isolated signal model. A positive value indicates that the overlapping signals model is preferred.\label{tab:logbf}}
\end{table}

\begin{figure*}
    \centering
    \includegraphics[width=\linewidth]{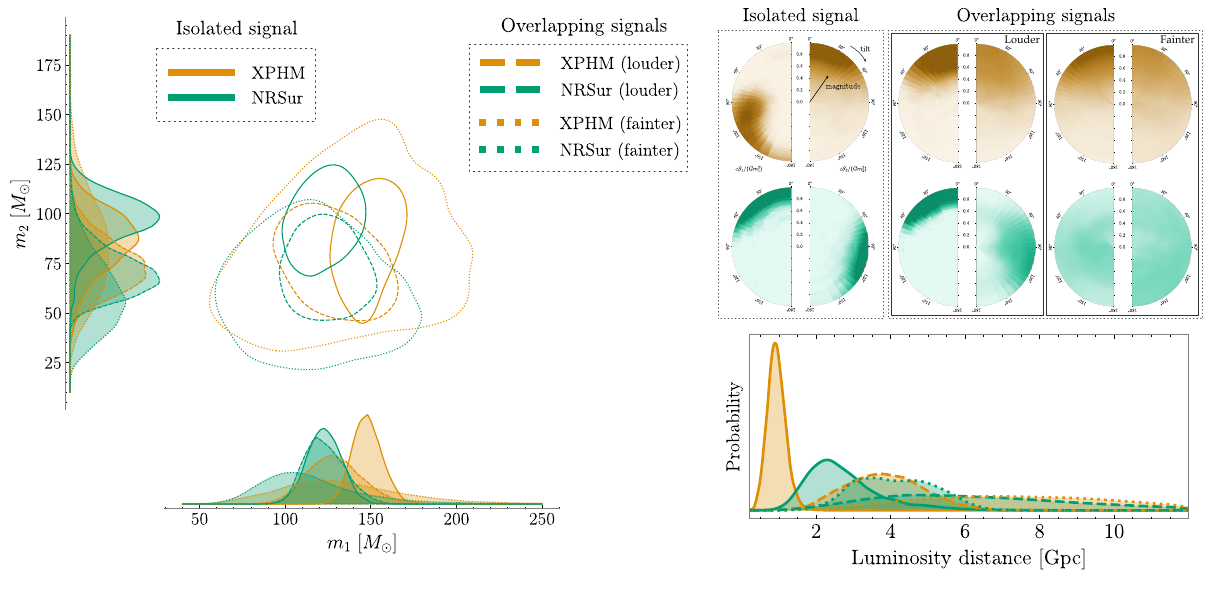}
    \caption{Comparison of the estimated source properties when using the isolated signal and the overlapping signals model. We show component masses in the left panel, component spins in top right panel, luminosity distance in the bottom right panel. We label the two signals recovered by the overlapping signals model by their SNRs, i.e., louder (light continuous lines) and fainter (light dashed lines). We compare the XPHM (brown) and the NRSur (green) analyses. When using the isolated signal model, we find discrepancies in the measurements between XPHM and NRSur analyses. These discrepancies are largely mitigated when using the overlapping signals model, e.g., compare the louder signal's source properties between XPHM and NRSur analyses. Similarly, the posterior distributions for the fainter signals agree between the two waveforms, which are significantly broader due to the low SNR of the signal.}
    \label{fig:properties_all_in_one}
\end{figure*}

\section{\label{sec:gw231123} One isolated signal versus two overlapping signals}We refer to the model with one isolated signal by $\mathcal{S}$ and two overlapping signals by $\mathcal{OS}$. The $\log$ Bayes factor comparing the two models is given by 
\begin{equation}
    \label{eq:bfactor}
    \lbf = \log_{10}{\mathcal{Z}_\mathcal{OS}} - \log_{10}{\mathcal{Z}_\mathcal{S}}.
\end{equation}
A positive $\lbf$ indicates that the two overlapping signals model is preferred over the isolated signal model. 

We compute the $\bayesf$ for four waveform models: IMRPhenomXPHM-SpinTaylor~\citep{Pratten:2020ceb, Colleoni:2024knd}, NRSur7dq4~\citep{Varma:2019csw}, IMRPhenomTPHM~\citep{Estelles:2021gvs}, and IMRPhenomXO4a~\citep{Thompson:2023ase}. Hereafter, for brevity, we refer to them by XPHM, NRSur, TPHM, and XO4a, respectively. We find that the overlapping signals model is preferred over the isolated signal model (Table~\ref{tab:logbf}) for all four waveforms. We note that the overlapping signals model has twice the number of model parameters compared to the isolated signal model. According to Occam's razor~\citep{Thrane:2018qnx}, if two models fit the data equally well, the Bayesian model selection will prefer the model with fewer parameters, penalising the model with more parameters. However, the overlapping signals model overcomes this penalty and is preferred over the isolated signal model. 

Comparing the XPHM and NRSur analyses, we find that measurements of masses, distance, and spins are significantly different when using the isolated signal model. These discrepancies are resolved when using the overlapping signals model (Figure~\ref{fig:properties_all_in_one}). Specifically, when using the overlapping signals model, the measurement of the masses, spins, and luminosity distance becomes consistent between the XPHM and NRSur analyses for the louder signals. {This does not provide direct proof of the overlapping-signal hypothesis, i.e. it is not a sufficient condition. 
Rather, it may be viewed as a necessary consequence of that hypothesis, and provides an alternative to waveform systematics as the cause of the posterior discrepancies in the single-signal model. } For the fainter signals, the low signal-to-noise-ratio (SNR) leads to significantly broader posterior distributions, though they still agree well between the two models. We find that the parameter estimates of the fainter signal, including its masses and sky location, overlap with those of the louder one, suggesting that the two signals could share the same source properties. We also find the difference in the arrival time to be 20 milliseconds between the two signals of the overlapping signals model. These findings are consistent with the lensing analyses of the event, where one typically measures the arrival time difference between two lensed images~\citep{xikai:2025, goyalab:2025, Liuab:2025}. 

{While the component masses are broadly consistent between the overlapping-signal and single-signal models, the inferred spin magnitudes tend to be lower in the overlapping-signal case. Since high spins form the basis of many interpretations of the astrophysical origin of GW231123~\citep{Stegmann:2025cja, Li:2025fnf, Li:2025pyo, Bartos:2025pkv, Delfavero:2025lup}, adopting the overlapping-signal results may reduce the statistical significance of those interpretations. In addition, waveform uncertainties are generally smaller in the low-spin region of parameter space, which may help to explain the reduced posterior discrepancies between XPHM and NRSur. }

Comparing the TPHM and XO4a analyses, we do not find any significant difference in the measurement of source properties, whether we use the isolated or overlapping signal model. This is consistent with the $\bayesf$ values for XO4a and TPHM analyses, which are much smaller compared to the ones obtained from XPHM and NRSur analyses. 

\section{\label{sec:wfsys}Presence of waveform systematics}The difference in $\bayesf$ values between different waveforms may be attributed to waveform systematics. Among all four waveforms, the TPHM and XO4a waveforms show stronger evidence in favour of the isolated signal model for GW231123 (first row of Table~\ref{tab:logbf}). If we assume that there is an additional signal present in the data overlapping with GW231123, the values in Table~\ref{tab:logbf} imply that the XO4a and TPHM waveforms are more susceptible when fitting to a distorted signal, compared to the other two waveforms. 

To verify this, we simulate two overlapping signals in zero noise and recover them with the isolated signal model. We perform four analyses: the simulated data consists of two overlapping signals generated by the NRSur waveform, which are then recovered with the isolated signal model using each of the four waveform models separately. We select the maximum likelihood parameters of the NRSur analysis of GW231123 with the overlapping signal model as injection parameters.

\begin{figure}
    \includegraphics[width=\linewidth]{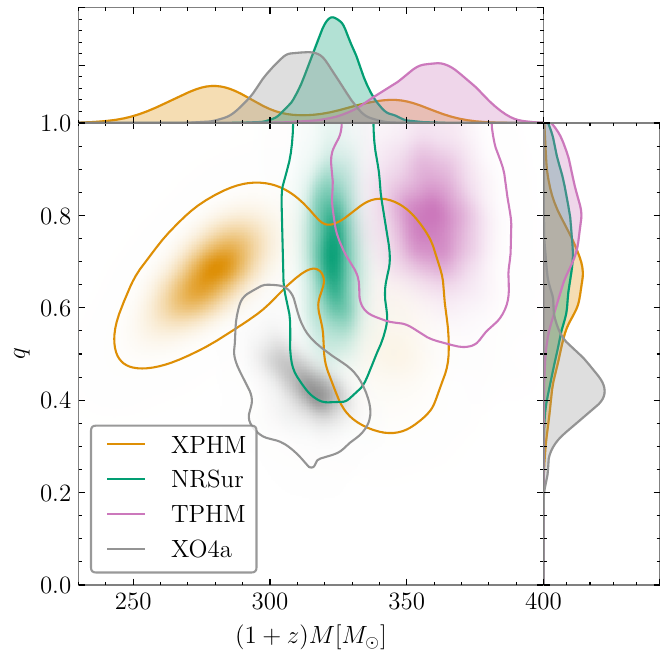}
    \vspace{1em}
    \hspace{1.5em}\includegraphics[width=0.95\linewidth]{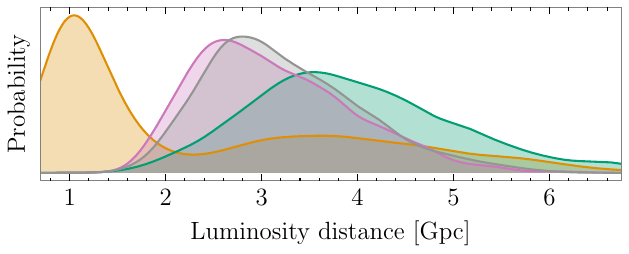}
    \caption{Posterior distributions when we recover two simulated overlapping signals using the isolated signal model with four different waveforms. The top panel shows the measurements of detector frame total mass and mass ratio; with the contours showing 90\% confidence interval. The bottom panel show the measurements of luminosity distance. With these simulations, we are able to reproduce the discrepancies in the estimated source properties between different waveforms reported in the GW231123 data release (c.f. Figures 7 and 8 in Ref.~\citep{LIGOScientific:2025rsn}).}
    \label{fig:os-sys}
\end{figure}

These simulations corroborate the results of the GW231123 analysis (Table~\ref{tab:zero-noise-z}). Specifically, we find that when we attempt to recover two overlapping signals with the isolated signal model, the XO4a waveform gives the largest evidence, followed by TPHM, NRSur, and XPHM. Besides the evidence values, we show the estimated source properties from these simulations in Figure~\ref{fig:os-sys}. Specifically, we show the detector frame total mass, mass ratio, and the luminosity distance. We observe discrepancies in the measurements between different waveforms, which are similar to the ones observed for GW231123~\citep{LIGOScientific:2025rsn}, indicating that the presence of an additional signal could cause this behaviour.

\begin{table}[H]\centering
\renewcommand{\arraystretch}{1.2}
\begin{tabular}{|c|c|c|c|c|}
\hline
             & \textbf{XPHM} & \textbf{NRSur} & \textbf{TPHM} & \textbf{XO4a} \\ \hline
        $\log_{10} \mathcal{Z}_{\mathcal{S}}$ & 171.08 & 177.81 & 179.60 & 180.21\\ \hline
\end{tabular}
\caption{Log evidence when we recover two simulated overlapping signals with the isolated signal model in zero noise.\label{tab:zero-noise-z}}
\end{table}

\begin{figure}
    \centering
    \includegraphics[width=\linewidth]{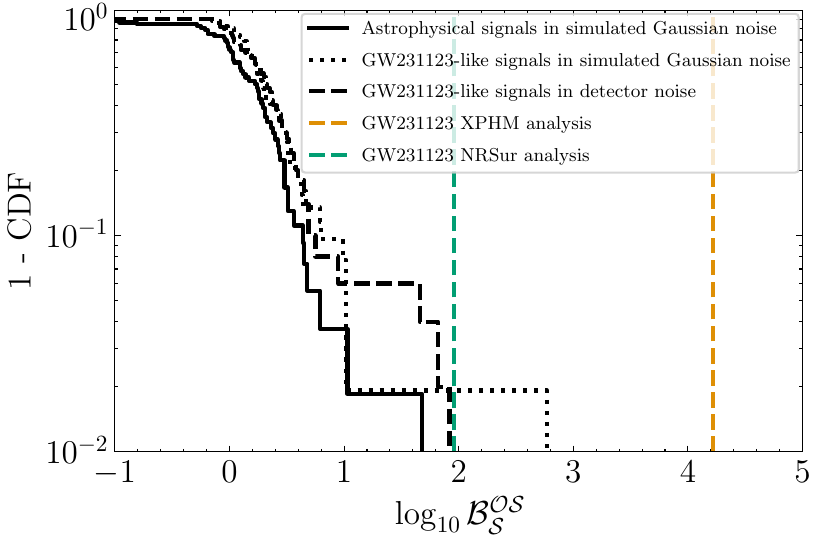}
    \caption{Complementary cumulative distributions (1-CDF) of the background statistic corresponding to the foreground ($\lbf$). The background statistic (black lines) represent how often the isolated BBH signals mimic the features of a pair of overlapping signals. The dashed green (brown) lines show the foreground statistic for NRSur (XPHM) analyses, i.e., $\lbf$ obtained by analysing the GW231123 signal. The background distribution in detector noise is generally above the background of the simulated Gaussian noise, indicating noise artifacts play may mimic the signs of overlapping signals. The distance between the dashed brown and green line may indicate the presence of waveform systematics.}
    \label{fig:background_hist}
\end{figure}

\section{\label{sec4}Estimating the background statistic}To estimate the significance of $\lbf$ comparing the overlapping and the isolated signal models, we estimate the corresponding background statistic. Specifically, we want to find out how often an isolated BBH signal could imitate the features of two overlapping signals due to the presence of noise fluctuations. To that end, we simulate a set of 100 BBH signals; 50 of which are chosen from the up-to-date BBH population distribution~\citep{LIGOScientific:2025pvj}, and the remaining 50 are chosen from the posterior distribution of GW231123~\citep{ligo_scientific_collaboration_2025_16004263}. The latter subset is to ensure we cover the parameter space of high total mass and high spin BBHs. To compute the background statistic, we perform parameter estimation on these isolated, simulated BBH signals using the isolated signal and the overlapping signals models.

When estimating background using GW231123-like signals, we rely on NRSur waveform for injection and recovery. We consider GW231123-like signals in simulated Gaussian noise as well as detector noise around GW231123 but excluding the event. The latter choice is to estimate how the noise artifact may falsely indicate the presence of overlapping signals. When estimating the background using astrophysical BBH signals, we rely on XPHM waveform for injection and recovery due to limited parameter-space coverage of NRSur~\citep{LIGOScientific:2025slb, Varma:2019csw}. For the astrophysical background, we only consider simulated Gaussian noise. The case of astrophysical BBH in detector noise should give a conservative estimate of the background compared to the GW231123-like signals in detector noise so we do not consider it here. We present the collective results of the background estimation in Figure~\ref{fig:background_hist}. 

In Figure~\ref{fig:background_hist}, the green (brown) line shows the foreground $\lbf$ for NRSur (XPHM) analyses. The black lines show the background statistic for events drawn from an astrophysical population (solid) and GW231123-like events (dashed and dotted lines for real detector and Gaussian noise, respectively). 
We find that all of the astrophysical background is below the foreground statistic of the XPHM analysis (brown), which indicates strong support for the overlapping signals model when using the XPHM waveform. Further, the large difference between the brown and green lines may be caused by waveform systematics between NRSur and XPHM. The background estimated using the detector noise is consistently above that of simulated Gaussian noise, indicating that the noise artefacts play an important role when comparing the two models, as seen for our GW231123-like injections. Additionally, we find that the GW231123-like signals are more likely to falsely imitate the signs of overlapping signals than the broader BBH population.

{Although background studies usually focus on false positives caused by noise fluctuations, we note that waveform systematics can also produce false positives in overlapping-signal analyses, as waveform errors may be absorbed by an additional signal. This may be particularly relevant for the high-mass scenario of GW231123, since the signal is dominated by the merger stage, where waveform uncertainties are largest. As pointed out by \citet{LIGOScientific:2025rsn}, a more accurate waveform model would likely be required before drawing confident conclusions for GW231123.}

\section{GW190521}Similarly, we analyse GW190521~\citep{LIGOScientific:2020iuh}, the heaviest BBH system prior to the detection of GW231123, using the isolated signal and the two overlapping signals model. We obtain $\lbf$ of 0.27 for XPHM, $-0.02$ for NRSur, $0.56$ for TPHM, and $0.54$ for XO4a waveform. The values of $\bayesf$ across different waveforms are more comparable, indicating the effect of waveform systematics may be negligible for GW190521. We also note that the inferred luminosity distance of the fainter signal consistently shifts toward higher values, which, in combination with the Bayes factors, disfavour the presence of an additional signal for this event. 

\section{\label{sec3}Prior odds of two overlapping signals}Although the Bayesian model selection favours the overlapping signals model, the prior odds of such an occurrence are expected to be low~\citep{Relton:2021cax}. In addition, both signals in the overlapping model lie in the high-mass tail of the astrophysical compact binary population~\citep{LIGOScientific:2025pvj}, further reducing the prior odds. To get an estimate of the prior odds, we adopt the high-mass BBH merger rate inferred from the observation of GW190521, $\mathcal{R} = 0.13^{+0.30}_{-0.11} \mathrm{Gpc}^{-3} \mathrm{yr}^{-1}$, and assume this merger rate is constant across the comoving volume. Given our prior bounds ($z_\mathrm{max}=1.6$ and $\Delta T = 0.4$s), the probability of observing another high-mass BBH event within a time window $\Delta T$ when one such signal is already present can be expressed using Poisson statistics as~\citep{Relton:2021cax}
\begin{equation}
    p(\mathcal{OS}|\mathcal{S}) = 1- \exp \left( -\Delta T\int_0^{z_\mathrm{max}} \frac{\mathcal{R}dV_c}{(1+z)dz}dz \right),
\end{equation}
which yields $p(\mathcal{OS}|\mathcal{S}) \sim 10^{-7}$. Multiplying by the Bayes factor, we obtain the final posterior odds ratio of observing two overlapping BBH signals to be $\lesssim  10^{-3}$, an exceedingly small number. {Since the overlapping signal model accommodates a number of phenomena, the low prior odds of two overlapping signals implies the other ones, such as gravitational lensing or overlapping glitches, would be sensible alternatives. }

\section{\label{sec5}Conclusions and discussions}We perform Bayesian model selection analyses on GW231123 to determine which model, between the overlapping signals and the isolated signal, is favoured by the data. For all waveform models used for the analysis, we find that the flexible signal model allowing for two overlapping gravitational wave signals is favoured over the isolated signal model when analysing GW231123. The overlapping signals model largely mitigates the discrepancies in measurements of mass, distance, and spins between different waveforms. In an additional set of simulations, we are able to reproduce the discrepancies between different waveform models similar to GW231123, when neglecting the presence of the overlapping signals. 

{However, interpreting high-mass binary black hole systems is challenging because only the last few cycles of the signal are observed, and a variety of models may fit the data, thereby affecting the inferred source properties. We cannot confidently establish an overlapping-signal interpretation for GW231123, partly because noise features may mimic signatures of overlapping signals, as indicated by the background estimates, and partly because of waveform systematics, as indicated by the comparison between different waveform analyses. That said, our findings on GW231123 remain valuable, as this is the first GW event for which the overlapping-signal model provides a better fit than the single-signal model; no comparable behaviour is seen in another similarly high-mass BBH event, GW190521. }

{While the detection of two overlapping gravitational wave signals is disfavoured by the current detection rate estimates, the overlapping signal model could be viewed as a superset of other potential phenomena which could be detected with current detector sensitivities. One possibility is the presence of faint noise artifacts near the signal~\citep{Ray:2025rtt}, although similar noise transients occurring coherently in both detectors are not commonly observed. The overlapping-signal model can also describe the case of gravitationally lensed signals in which two images overlap in time due to a small arrival-time delay. This degeneracy between lensing and overlapping signals makes it important to confidently rule out the overlapping-signal scenario. In this work, we find that the overlapping signal interpretation is disfavoured by astrophysical rate estimates, while the inferred source properties remain consistent with a possible lensing interpretation. These results therefore motivate further investigation of the gravitational-lensing scenario for GW231123.}

{As the detector sensitivity increases and as we transition towards the era of third-generation gravitational wave detectors~\citep{Abac:2025saz, reitze2019_CosmicExplorerContribution}, it will become possible to observe subtle effects from both lensing and overlapping signals.
In such scenarios, it becomes crucial to develop a methodology that can distinguish between the two effects. Our work takes a first step in that direction. } 


\section{Acknowledgments}The authors would like to thank Jonathan Gair, Yifan Wang, Zhen Pan, and the LIGO--Virgo--KAGRA parameter estimation group for helpful discussions and suggestions. The authors are grateful for the computing resources provided by Cardiff University funded by STFC grant ST/I006285/1 and ST/V005618/1, and by the Consortium
des Équipements de Calcul Intensif (CÉCI) funded by the Fonds de la Recherche Scientifique de Belgique
(F.R.S.-FNRS) under Grant No. 2.5020.11 and by the Walloon Region. QH and JV are supported by STFC grant ST/Y004256/1. JH is a FRIA grantee of the Fonds de la Recherche Scientifique - FNRS. HN, MW, and CVDB are supported by the research programme of the Nederlandse Organisatie voor Wetenschappelijk Onderzoek (Netherlands Organisation for Scientific Research, NWO). HN and CVDB further acknowledge the support of NWO through the grant number OCENW.XL21.XL21.038. We also acknowledge the use of the computing infrastructure of Nikhef, which is co-supported by the NWO. This material is based upon work supported by NSF's LIGO Laboratory which is a major facility fully funded by the National Science Foundation.



\bibliography{refs.bib}



\end{document}